\documentclass{article}
\usepackage{amssymb}
\usepackage{amsfonts}
\usepackage{graphicx}

\title{Excitation Spectrum at the Yang-Lee Edge Singularity of 2D Ising
 Model on the Strip}

\author{
\textrm{SMAIN BALASKA}\\%
\textit{Laboratoire de PhysiqueTh\'{e}orique  d'Oran, D\'{e}partement
 de Physique}  \\
\textit{Universit\'{e} d'Oran Es-s\'{e}nia, 31110 Es-s\'{e}nia, Algeria
 }\\
sbalaska@yahoo.com; balaska.smain@univ-oran.dz\\
\\
\textrm{JOHN F. MCCABE}\\
\textit{412 Morris Avenue \# 34, Summit, New Jersey 07901, USA}\\
jfmccabe2@earthlink.net\\
\\
\textrm{TOMASZ WYDRO} \\
\textit{Laboratoire de Physique Mol\'{e}culaire et des Collisions} \\
\textit{Universit\'{e} Paul Verlaine-Metz, 1 bvd Arago, 57078 Metz,
 France} \\
wydro@univ-metz.fr
}

\begin{document}

\maketitle

\begin{abstract}
At the Yang-Lee edge singularity, finite-size scaling behavior is
used to measure the low-lying excitation spectrum of the Ising
quantum spin chain for free boundary conditions. The measured
spectrum is used to identify the CFT that describes the Yang-Lee
edge singularity of the 2D Ising model for free boundary conditions.
\end{abstract}


\begin{center}
\bigskip
\end{center}

In 1985, Cardy\cite{IsingYLS} identified the $(A_{4}, A_{1})$
minimal conformal field theory (CFT)\cite{BPZ,Friedan,ADE1} with the
Yang-Lee edge singularity\cite{YL} of the 2-dimensional (2D) Ising
model on the plane. Based on this identification, Cardy predicted
properties of the Yang-Lee edge singularity.\cite{IsingYLS} Several
of these CFT predictions have been confirmed numerically using the
2D Ising lattice model on the cylinder.\cite{Zuber,Uzelac,MC}

The  2D Ising model can also be defined on surfaces
having boundaries.  On such surfaces, CFT on the 1/2-plane should
describe critical properties,\cite{bd1} e.g., at the Yang-Lee edge
singularity. In general, different boundary conditions on the
1/2-plane will define different CFTs.\cite{bd1}  Each of these CFTs
has a spectrum corresponding to one or more of the Verma modules of
the corresponding CFT on the plane.\cite{bd2}

The present article is organized as follows.  First, it presents
tools that were used to study the Yang-Lee edge singularity of the
2D Ising model.  The tools include quantum spin chains and
finite-size scaling.  Second, it presents the excitation spectrum
of candidate CFTs for the Yang-Lee edge singularity of the Ising
model with free boundary conditions.  Third, it presents
finite-size measurements of the low-lying excitation spectrum at
the Yang-Lee edge singularity of the Ising quantum spin chain with
free boundary conditions.  Fourth, it compares the measured
spectrum to the spectra of the candidate CFTs.

To measure the excitation spectra, the extreme anisotropic limit is
taken of the 2D Ising model.  In this limit, each critical point of
a 2D model becomes a critical point of a corresponding quantum spin
chain.\cite{Fradk78} In particular, each critical point of the 2D
Ising model with free boundary conditions on the infinite strip will
correspond to a critical point of the Ising quantum spin chain with
free boundary conditions at its ends. The Yang-Lee edge
singularities of these two models will have the same critical
properties.

The Ising quantum spin chain for the 2D Ising model has a
Hamiltonian given by:\cite{Gehlen87}
\begin{equation}
H_{Ising}={\sum_{n=1}}^{N}\{-t\sigma_{z}(n)\sigma_{z}(n+1)-h\sigma_{z}(n)+\sigma_{x}(n)\}.
\label{HIsing}
\end{equation}
In eq. (\ref{HIsing}), $N$ is the number of sites, $\sigma_{x}(n)$
and $\sigma_{z}(n)$ are 2x2 Pauli spin matrices at site "$n$", "$h$"
is an external magnetic field, and "$t$" is a ferromagnetic
spin-spin coupling, i.e., $t > 0$.

At the Yang-Lee edge singularity, the external magnetic field, $h$,
is purely imaginary, i.e., $h = iB$ for a real $B$. On a quantum
spin chain of length, $N$, the phenomenological renormalization
group (PRG) defines special values, $iB_{YL}(N)$, of the magnetic
field. For a length of $N$, the special value $iB_{YL}(N)$ satisfies
the PRG equation:\cite{Derrida,Zuber}
\begin{equation}
[N -1]m(iB_{YL}(N), N-1)= [N]m(iB_{YL}(N), N) . \label{PRG}
\end{equation}
In eq. (\ref{PRG}), $m(iB, N)$is the energy gap, which is equal to
$\left[E_{1}(iB, N) - E_{0}(iB, N)\right]$.  The energies
$E_{0}(iB, N)$ and $E_{1}(iB,N)$ are the energies of the
respective ground state "0" and first excited state "1" for the
Ising quantum spin chain of length N. As $N\to\infty$, the special
values, $iB_{YL}(N)$, converge to the Yang-Lee edge singularity.
The scaling behavior of physical quantities at solutions of the
PRG equation provides the scaling behavior of same physical
quantities near a corresponding critical
point.\cite{Derrida,Zuber}

For a spin-spin coupling t of 0.1, Table [\ref{tab:BC}] shows our
measurements of the special values of the magnetic field that solve
the PRG eq. (\ref{PRG}).
\begin{table}[h]
\centering
\begin{tabular} {|c|c|}
\hline
$N$& $B_{YL}(N)$\\
\hline
 6 & 0.64452828  \\
\hline
 7 & 0.64155254  \\
\hline
 8 & 0.63986508  \\
\hline
 9 & 0.63884984  \\
\hline
 10& 0.63820860  \\
\hline
 11& 0.63778681  \\
\hline
 12& 0.63749969  \\
\hline
$\infty$& ---  \\
\hline
\end{tabular}
\caption{PRG values of $B_{YL}(N)$ when $t = 0.1.$ \label{tab:BC}}
\end{table}
The special values of the magnetic field of Table [\ref{tab:BC}]
correspond to the Yang-Lee edge singularity when free boundary
conditions are imposed on the Ising quantum spin chain. When
evaluated at the $B_{YL}(N)$'s, physical properties have finite-size
scaling behaviors corresponding to the Yang-Lee edge singularity in
the limit where $N \to\infty$.

In the same limit, CFT predicts that the leading scaling behavior of
the actual excitation energies with the length, N, of a quantum spin
chain will be:\cite{Cardy-scaling}
\begin{equation}
E_{i}(N)-E_{0}(N)=\xi 2\pi {\Delta_{i}/ N}. \label{gaps scaling}
\end{equation}
Here, $\Delta_{i}$ is the conformal dimension of the field "i", and
$\xi$ is non-universal constant that depends on the normalization of
the Hamiltonian for the Ising quantum spin chain. In eq. (\ref{gaps
scaling}), the excitation energies depend only on one conformal
weight, because physical boundaries reduce the symmetry of the CFT
to a single Virasoro algebra.

On the 1/2-plane, a CFT has a partition function that is a linear
combination of the Virasoro characters appearing in the partition
function of the same CFT on the plane.\cite{bd1,bd2} On the plane,
the $(A_{4}, A_{1})$ minimal CFT describes the Yang-Lee edge
singularity of the 2D Ising model. The partition function of the
$(A_{4}, A_{1})$ minimal CFT is constructed from (c = -22/5,
$\Delta$ = 0) and (c = -22/5, $\Delta$ = -1/5) Verma modules.
Thus, the CFT of the Yang-Lee edge singularity on the infinite
strip should have the states of the (c = -22/5, $\Delta$ = 0)
Verma module and/or the states of the (c = -22/5, $\Delta$ = -1/5)
Verma module. That is, there are three types of candidate CFT for
the Yang-Lee edge singularity of the 2D Ising model on the
infinite strip with free boundary conditions. The candidate CFTs
have the states of the (c = -22/5, $\Delta$ = 0) Verma module, the
states of the (c = -22/5, $\Delta$ = -1/5) Verma module, or the
states of both these Verma modules. For each of these candidate
CFT, Table [\ref{tab:CFT-SPECTRA}] shows the low-lying excitation
spectrum.
\begin{table}[h]
\centering
\begin{tabular} {|c|c|c|c|c|c|c|}
\hline
\multicolumn {7}{|c|}
{\bf Strip Boundary Conditions} \\
\hline
\multicolumn {7}{|l|}
{\bf CFT of (-22/5, 0) Verma module} \\
\hline
Excitation Energy & 2 & 3 & 4 & 5 & 6 & 7 \\
\hline
Degeneracy        & 1 & 1 & 1 & 1 & 2 & 2 \\
\hline
\multicolumn {7}{|l|}
{\bf CFT of (-22/5, -1/5) Verma module} \\
\hline
Excitation Energy & 2 & 4 & 6 & 8 & 10 & 12  \\
\hline
Degeneracy        & 1 & 1 & 1 & 2 & 2 & 3  \\
\hline
\multicolumn {7}{|l|}
{\bf CFT of (-22/5, 0) and (-22/5, -1/5) modules} \\
\hline
Excitation Energy & 2 & 10 & 20& 22& 30& 32  \\
\hline
Degeneracy        &   &   &   &   &   &  \\
\hline \multicolumn {7}{|c|}
{\bf $(A_{4}, A_{1})$ CFT on $\infty$-long cylinder} \\
\hline
Excitation Energy & 2 & 5& 10 & 12 & 15 & \\
\hline
Degeneracy        & 1 & 2 & 3 & 2  & 4 & \\
\hline
\end{tabular}
\caption{Low-Lying Excitation spectra of candidate CFTs
 \label{tab:CFT-SPECTRA}}
\end{table}
In Table [\ref{tab:CFT-SPECTRA}], the CFT predictions for the
excitation energies are normalized so that the lowest excitation
energy is two. This normalization removes both the dependency on
the non-universal constant $\xi$ and the dependence on the quantum
spin chain's length, N.\footnote{For this choice of normalization,
the measured normalized excitation energies will also turn out to
be actual conformal dimensions for this specific model.} For the
types of CFT that combine both Verma modules, Table
[\ref{tab:CFT-SPECTRA}] does not list state degeneracies, because
these degeneracies will depend on the number of copies of each of
the two Verma modules in the CFT. Table [\ref{tab:CFT-SPECTRA}]
also lists the low-lying excitation spectrum of the $(A_{4},
A_{1})$ minimal CFT, which describes the Yang-Lee edge singularity
on the infinitely long cylinder.  From Table
[\ref{tab:CFT-SPECTRA}], one also sees that the low-lying
excitation spectra very substantially distinguish the candidate
CFTs for the Yang-Lee edge from each other.

\begin{table}[h]
\centering
\begin{tabular} {|c|c|c|c|c|c|c|c|c|c|}
\hline
$N$   &  6    &  7     &  8     &  9     &  10    &  11    &  12  
  &$\infty$ \\
\hline
1st excitation &2.00&2.00 &2.00    &2.00    &2.00    &2.00    &2.00  
  & 2.00    \\
\hline
A       &3.10970 &3.08865 &3.07468 &3.06481 &3.05748 &3.05183 &3.04732
 & 3.02   \\
\hline
B       &4.01911 &4.01608 &4.01552 &4.01606 &4.01700 &4.01801 &4.01895
 & 4.02   \\
\hline
C       &4.72083 &4.77954 &4.82010 &4.85112 &4.87569 &4.89555 &4.91182
 & 5.05   \\
\hline
D1      &  ---   &5.35658 &5.47758 &5.56177 &5.62651 &5.67802 &5.71994
 & 6.01   \\
\hline
D2      &6.15534 &6.12718 &6.11081 &6.09947 &6.09136 &6.08516 &6.08017
 & 6.06   \\
\hline
E1      &  ---   &  ---   &5.95864 &6.13662 &6.26129 &6.35878 &6.43745
 & 6.90   \\
\hline
E2      &7.03192 &7.02799 &7.02804 &7.03040 &7.03483 &7.03688 &7.03914
 & 7.03   \\
\hline
\end{tabular}
\caption{Measured Low-lying excitation energies as a function of
length, $N$, of the Ising quantum spin chain. \label{tab:SPECTRA}}
\end{table}

\begin{figure} [h] \centering
\centering
    \includegraphics[scale=1.3]{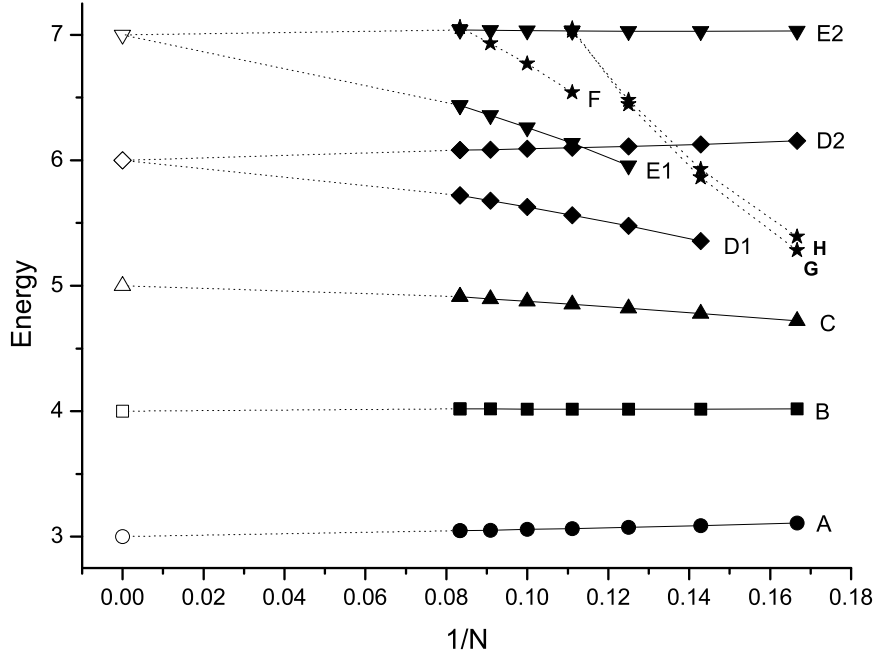}
    \caption{Measured Low-Lying Normalized
Excitation Energies for $N$ between 6 and
12.}\label{fig:EnergyLevels}
 \end{figure}

For special magnetic field values and spin-spin coupling of Table
[\ref{tab:BC}], the low-lying excitation spectrum measured for
Ising quantum spin chains of lengths between 6 and 12 sites is
given in Table [\ref{tab:SPECTRA}] and plotted in Figure
(\ref{fig:EnergyLevels}). The measured excitation energies have
been normalized by division by 1/2 times the lowest measured
excitation energy.\footnote{Figure (1) does not show the lowest
excitation energy, which is exactly two in this normalization.}
This normalization enables a direct comparison between the
measured excitation spectra of Table [\ref{tab:SPECTRA}] with the
spectra of the candidate CFTs as given in Table
[\ref{tab:CFT-SPECTRA}].

To compare the measured spectra of Fig.(\ref{fig:EnergyLevels}) to the
 predicted spectra
of Table [\ref{tab:CFT-SPECTRA}], it is necessary to determine the
limit of the normalized excitation energies of the states as the
quantum spin chain's length, $N$, becomes large. To find these limit
values, one must identify corresponding states in Ising quantum spin
chains of different length, $N$. In Fig.(\ref{fig:EnergyLevels}), our
 conclusions about
these correspondences are indicated by lines between measured
excitation energies for different lengths, $N$. The correspondences
were found by assuming a smooth scaling behavior and by using the
fact that level crossings should disappear as $N$ grows.

 In Fig.(\ref{fig:EnergyLevels}), it is
easy to identify sequences A, B, C, D1, D2, E1, E2, F, G, and
H\footnote{No state has been double counted in identifying the
sequences A - H.}  of normalized excitation energies. As expected,
crossings between different ones of these levels disappear as $N$
increases. The normalized excitation energies of the lowest
states, which are labeled by A, B, C, D1, D2, E1, and E2, smoothly
scale toward values of about 3 to 7 as $N \to\infty$. The
normalized excitation energies of the states labeled by F, G, and
H smoothly scale towards higher values as $N \to\infty$. For the
states F, G, and H, the evaluation of the limit of these higher
energy excitations as $N \to\infty$ is outside of our measurements
on Ising quantum spin chains of lengths between 6 and 12.

The last column of Table [\ref{tab:SPECTRA}] also shows normalized
excitation energies obtained by extrapolating the measured values
to the limit where $N \to\infty$. Each such extrapolation was made
by fitting normalized measured excitation energies of a
corresponding state, i.e., $E_i(N) - E_0(N)$, to an equation of
the following form:
\begin{equation}{
E_i(N) - E_0(N) = [E_i(\infty) - E_0(\infty)] + A_iN^{p_i} \ {\rm
with } \ p_i < 0.  \label{SCALING}}
\end{equation}
In eq. (\ref{SCALING}), $[E_i(\infty) - E_0(\infty)]$ is the limit
of the $i$-th excitation energy as $N \to\infty$, and $A_iN^{p_i}$
is a correction term whose form is motivated by finite-size scaling
considerations.

From the above-described extrapolations, we find that the
low-lying excitation spectrum for the Ising quantum spin chain at
the Yang-Lee edge singularity has the following sequence of
normalized excitation energies (degeneracies): 2(1), 3(1), 4(1),
5(1), 6(2), and 7(2). By comparing these results with the CFT
predictions of Table [\ref{tab:CFT-SPECTRA}], it is readily seen
that the measured low-lying excitation spectrum is the same as the
low-lying excitation spectrum of the minimal CFT that is based on
the (c, $\Delta$) = (-22/5, 0) Verma module.

In conclusion, the measured low-lying excitation spectrum of the
Ising quantum spin chain with free boundary conditions at the
Yang-Lee edge singularity is in excellent agreement with the
low-lying spectrum of the (c, $\Delta$) = (-22/5, 0) Verma module.
Thus, the (c, $\Delta$) = (-22/5, 0) Verma module defines the CFT
for the Yang-Lee edge singularity of the 2D Ising model on the
infinite strip with free boundary conditions.

\end{document}